\documentclass[aps,prd,twocolumn,superscriptaddress,nofootinbib,eqsecnum,showpacs]{revtex4}

\usepackage{bm}
\usepackage{amsmath}
\usepackage{amsfonts}
\usepackage{amssymb}
\usepackage{graphicx}

\usepackage[usenames, dvipsnames]{color}
\usepackage[colorlinks, pdfborder={0 0 0}, plainpages=false]{hyperref}

\usepackage{breakurl}
\usepackage{xspace} 
\usepackage{dcolumn}
\usepackage{multirow} 

\definecolor{CiteColor}{rgb}{0, 0.5, 0}
\hypersetup{citecolor=CiteColor}
\definecolor{RefColor}{rgb}{0.55, 0, 0}
\hypersetup{linkcolor=RefColor}

\newcommand{\ud}{\mathrm{d}}
\newcommand{\beq}{\begin{equation}}
\newcommand{\eeq}{\end{equation}}

\begin{document}

\title{The complete non-spinning effective-one-body metric \\ at linear order in the mass ratio}

\author{Enrico Barausse}
\affiliation{Maryland Center for Fundamental Physics \& Joint Space-Science Institute, Department of Physics, University of Maryland, College Park, MD 20742}

\author{Alessandra Buonanno}
\affiliation{Maryland Center for Fundamental Physics \& Joint Space-Science Institute, Department of Physics, University of Maryland, College Park, MD 20742}
\affiliation{Radcliffe Institute for Advanced Study, Harvard University, \\ 8 Garden St., Cambridge, MA 02138}

\author{Alexandre Le Tiec}
\affiliation{Maryland Center for Fundamental Physics \& Joint Space-Science Institute, Department of Physics, University of Maryland, College Park, MD 20742}

\begin{abstract}
  Using the main result of a companion paper, in which the binding energy
  of a circular-orbit non-spinning compact binary system is computed at leading-order beyond the test-particle
  approximation, the exact expression of the effective-one-body (EOB)
  metric component $g^\text{eff}_{tt}$ is obtained through first order in
  the mass ratio. Combining these results with the recent
  gravitational self-force calculation of the periastron advance for
  circular orbits in the Schwarzschild geometry, the EOB metric
  component $g^\text{eff}_{rr}$ is also determined at linear order in the mass ratio. These results assume 
  that the mapping between the real and effective Hamiltonians 
  at the second and third post-Newtonian (PN) orders holds at all PN orders.
  Our findings also confirm the advantage of resumming the PN dynamics around the test-particle 
limit if the goal is to obtain a flexible model that can smoothly connect the 
test-mass and equal-mass limits.
\end{abstract}

\date{\today}


\maketitle

\section{Introduction}
\label{sec:intro}

Although the ``two-body problem'' cannot be solved analytically in the
general theory of relativity, at least two approximation methods can
be used to tackle it. The first one dates back to Einstein's 1915
calculation of the relativistic perihelion advance of Mercury's
orbit \cite{Ei.15}, and is based on a perturbative treatment
in powers of the ratio $v/c$ between the binary's relative velocity
$v$ (in the center-of-mass frame) and the vacuum speed of light
$c$. At the lowest order of approximation, this approach gives back
the well-known Newtonian solution, and is therefore dubbed
``post-Newtonian'' (PN) expansion; see \textit{e.g.} Ref.~\cite{Bl.06}
and references therein. Currently, the two-body dynamics of
  non-spinning compact objects is known through 3PN order\footnote{As usual we refer to $n$PN as the order corresponding to terms $\mathcal{O}(c^{-2n})$ with respect to the Newtonian acceleration in the equations of motion, or with respect to the quadrupole 
formula in the radiation field.} \cite{Da.al.01,Bl.al.04}, and the
  gravitational-wave fluxes of energy and angular momentum through 3.5PN \cite{Bl.al2.05,Ki.08,Bl.al.08} and 3PN \cite{Ar.al2.09} orders for circular 
  and eccentric orbits, respectively. Spin effects have also been computed in the dynamics
  and gravitational radiation \cite{Ki.95,Bl.al.06,Bl.al2.11} using both
  Hamiltonian \cite{Da.al.08,St.al.08,St.al2.08,HeSc2.08,HaSt.11}
  and Lagrangian \cite{Ki.95,Fa.al.06} formulations. The effective-field-theory approach applied to gravity \cite{GoRo.06} has
  confirmed some of these results \cite{FoSt.11,Po.al2.11}, and has pushed 
  the calculations to higher PN orders for spinning bodies \cite{Po.al2.11,PoRo.06,
  Po.06,PoRo.08,PoRo2.08,Po.10,Le.10,Le2.11}.
  
The second approximation method also dates back to 1915, namely to
Karl Schwarzschild's famous wartime calculation of the gravitational
field of a spherically symmetric body
\cite{Sc.16,Sc2.16}. While non-rotating
black holes are described by the Schwarzschild metric, rotating
black holes are represented by the Kerr solution \cite{Ke.63}.
The motion of test masses in the Schwarzschild or Kerr geometries is naturally
described by the geodesic equations, which are valid for arbitrarily
high values of the orbital velocity $v$ (\textit{i.e.} the geodesic equations formally
include all PN corrections). If the finite mass of the particle and
its backreaction on the background geometry are taken into account,
the orbits will deviate from geodesic motion under the effect of the
gravitational self-force (GSF) \cite{Po.al.11}.  More formally,
Schwarzschild or Kerr geodesics can be seen as the orbital motion of
the binary at the zeroth order of approximation in the mass ratio,
while at first order the two-body dynamics is regulated by the
so-called ``MiSaTaQuWa'' equation \cite{Mi.al.97,QuWa.97}. The
GSF can be split into a dissipative component related to
gravitational-wave emission --- which is described by the
Regge-Wheeler and Zerilli equations in a Schwarzschild background
\cite{ReWh.57,Ze3.70}, and by the Teukolsky equation in
Kerr \cite{Te.72,Te.73} ---, and a conservative component responsible
for secular effects such as the periastron advance \cite{De.08,BaSa.09,BaSa.11}.

A notable event in the history of the general relativistic two-body
problem took place almost a century after Einstein's and
Schwarzschild's early work, in 2005, with the first
numerical simulations of the inspiral, merger, and ringdown of a
system of two non-spinning black
holes \cite{Pr.05,Ca.al.06,Ba.al.06}. While these results
constitute a truly remarkable achievement, current ``state-of-the-art'' numerical
relativity (NR) simulations are still too time-consuming to provide gravitational waveforms covering the
whole parameter space of binary black-hole systems, especially for small mass ratios \cite{LoZl.11} and black holes 
with large spins \cite{Lo.al.11,Lo.al.12}.

A semi-analytical approach that is flexible enough to incorporate
information from both PN expansions and black-hole perturbations, as
well as from NR simulations, is the effective-one-body (EOB)
method \cite{BuDa.99}. The basic idea behind this
construction is to map the orbital dynamics of an arbitrary mass-ratio
compact binary system onto that of a test particle in a suitable
background spacetime. In order for the EOB model to have the correct
test-particle limit, this effective background metric,
$g^\text{eff}_{\alpha\beta}$, must clearly reduce to that of a
Schwarzschild black hole when one of the masses goes to zero (for non-spinning binaries). In
addition, such a mapping is known to exist for \textit{any} mass ratio
at the Newtonian level, because in Newtonian gravity one can always
map the dynamics of a binary system with masses $m_1$ and $m_2$ onto
the motion of an effective particle with reduced mass $\mu = m_1
m_2/(m_1+m_2)$ around a body with total mass $M=m_1+m_2$. It is
therefore natural to try to achieve the EOB mapping by considering an
effective particle with mass $\mu$ moving in a time-independent and
spherically symmetric \textit{deformed} Schwarzschild spacetime with
total mass $M$,\footnote{Throughout this paper we use geometrical unit $G = c = 1$.}
\beq\label{eff_metric}
	\ud s^2_\text{eff} = g^\text{eff}_{tt}(r;\nu) \, \ud t^2 + g^\text{eff}_{rr}(r;\nu) \, \ud r^2 + r^2 \ud \Omega^2 \, ,
\eeq
where the deformation is regulated by the symmetric mass ratio $\nu \equiv \mu/M$.
In the test-particle limit $\nu \to 0$, we recover (by construction) the Schwarzschild result
$g^\text{eff}_{tt} = - 1 / g^\text{eff}_{rr} = - 1 + 2M / r$. Beyond that limit, the $\nu$-dependence of the EOB potentials $g^\text{eff}_{tt}$ 
and $g^\text{eff}_{rr}$ encodes crucial information about the dynamics of the real binary system.

Buonanno and Damour \cite{BuDa.99} showed that, for spinless
binaries, the EOB mapping can be achieved not only at the Newtonian
level, but also at the 1PN and 2PN orders, obtaining the following relation
between the Hamiltonian $H_\mathrm{real}$ of the real binary system 
and the Hamiltonian $H_\mathrm{eff}$ of the effective particle:
\begin{equation}\label{mapping}
	H_\mathrm{real} = M\,\sqrt{1 + 2 \nu \left( \frac{H_{\rm eff}}{\mu} - 1 \right)} .
\end{equation}
Remarkably, this formula coincides with that introduced by
Br\'ezin, Itzykson and Zinn-Justin \cite{Br.al.70} in quantum electrodynamics to map
the one-body relativistic Balmer formula with the two-body
one; for example it can relate some
of the energy levels of positronium (an equal-mass system comprised
of an electron and an anti-electron, described by the real two-body
Hamiltonian $H_{\rm real}$) to those of the hydrogen atom (an
``extreme'' mass-ratio system described by the effective Hamiltonian
$H_{\rm eff}$).

Since the original paper \cite{BuDa.99}, the EOB mapping has been
extended to 3PN order for non-spinning binaries \cite{Da.al3.00}, and
shown to exist also for spinning binaries, through 3.5PN order in the
spin-orbit terms, and 2PN order in the spin-spin
terms \cite{Da.01,Da.al2.08,Ba.al.09,BaBu.10,BaBu.11,Na.11}. Furthermore,
the EOB construction has grown to include a model for the
gravitational waveforms \cite{BuDa.00,DaNa.07,Da.al.09,FuIy.10,Pa.al2.11}, allowing detailed comparisons (and
calibrations of the EOB model's unknown parameters) with NR waveforms
for non-spinning and spinning comparable-mass binaries \cite{Bu.al.07,Bu.al2.07,Pa.al.08,
Bo.al.08,Bu.al.09,Pa.al.10,DaNa.08,Da.al3.08,Da.al4.08,DaNa.09,Pa.al.11}, as well as with
Regge-Wheeler-Zerilli \cite{DaNa.07,Na.al.07,BeNa.10,Be.al.11,Be.al2.11} and
Teukolsky waveforms \cite{Pa.al2.11,Yu.al.10,Yu.al.11,Ba.al2.12}
for small and extreme mass-ratios.

More recently, information coming from GSF calculations
has started to be included into the EOB model. 
References \cite{Da.10,BaBu.10,Pa.al.11} used the frequency shift of
the Schwarzschild innermost stable circular orbit (ISCO) induced by the conservative GSF, as recently calculated by Barack and
Sago \cite{BaSa.09} (see also Ref.~\cite{BaSa.10}), to constrain some
unknown parameters entering the $g^\text{eff}_{tt}$ component of the
EOB effective metric \eqref{eff_metric} for spinless binaries, and regulating the unknown higher PN orders.
Ref.~\cite{Da.10} also suggested using GSF data to determine a certain combination $\rho_\text{SF}$ of the free functions parametrizing the $g^\text{eff}_{tt}$ and $g^\text{eff}_{rr}$ components of the EOB effective metric, at linear order in the mass ratio. That suggestion was then realized in Ref.~\cite{Ba.al.10}, which calculated the strong-field behavior of $\rho_\text{SF}$ using the GSF contribution to the periastron advance for quasi-circular orbits in a Schwarzschild background.

Besides the ISCO frequency shift and the periastron advance,
a third physically meaningful (\textit{i.e.} coordinate
invariant) perturbative result that has been obtained within the GSF framework 
is the conservative effect of the self-force on the ``redshift observable'' $z \equiv -
u_\alpha K^\alpha$, $u^\alpha$ being the particle's four-velocity and
$K^\alpha$ the helical Killing vector of the perturbed Schwarzschild
geometry \cite{De.08,Sa.al.08,Sh.al.11,Bl.al.10,Bl.al2.10}.
Refs.~\cite{De.08,Bl.al.10,Bl.al2.10} (see also \cite{Bl.al.11}) demonstrated a very good agreement between the numerical GSF result and the PN prediction.

The quantity $z$ measures the redshift of light rays emitted from the
particle, and received far away from the binary system, along the helical
symmetry axis perpendicular to the orbital plane
\cite{De.08}. In the companion paper \cite{Le.al3.11}, we put forward a different
interpretation for the redshift observable. Building on the first law
of binary point-particle mechanics recently established in
Ref.~\cite{Le.al2.11}, we relate $z$ to the binding energy of a
binary system of non-spinning compact objects through next-to-leading
order in the mass ratio. Using the numerical results of
Refs.~\cite{De.08,Sa.al.08,Sh.al.11,Bl.al.10,Bl.al2.10},
in which the GSF contribution to the redshift observable of a
non-spinning particle moving on a circular orbit around a
Schwarzschild black hole was calculated, we derive an explicit
expression for the binding energy.

In this paper, we use this new result for the binding energy, 
assuming that the mapping (\ref{mapping}) holds at all 
PN orders, to derive the \textit{exact} expression of 
the $g^\text{eff}_{tt}$ component of the EOB effective metric \eqref{eff_metric} for non-spinning binaries, through
\textit{linear} order in the mass ratio. The resulting expression exactly
reproduces the ISCO frequency shift calculated by Barack and Sago, and
goes far beyond the results of Refs. \cite{Da.10,BaBu.10,Pa.al.11}
that could only constrain certain combinations of unknown parameters
appearing in the EOB potential $g^\text{eff}_{tt}$.

Moreover, thanks to the exact
knowledge (at least through linear order in the mass ratio) of
$g^\text{eff}_{tt}$, we are able to use the GSF results of
Ref.~\cite{Da.10,Ba.al.10} for the periastron advance to
determine the EOB metric component $g^\text{eff}_{rr}$
\textit{exactly} (through linear order in the mass ratio) for
non-spinning binaries. This constitutes significant advance over
earlier results that could only constrain combinations of unknown functions entering $g^\text{eff}_{tt}$ and $g^\text{eff}_{rr}$. Our
results therefore completely determine the EOB effective metric \eqref{eff_metric} for a
system of non-spinning compact objects, at first order in the
mass ratio.

This paper is organized as follows. In Sec.~\ref{sec:EOB-SF-gen}, after briefly reviewing the 
EOB effective metric and Hamiltonian dynamics, we use the binding energy computed in Ref.~\cite{Le.al3.11} to derive the exact correction to $g^\text{eff}_{tt}$ that is linear in the 
mass ratio. Moreover, employing  the results of the periastron advance for circular orbits 
from Refs. \cite{Da.10,Ba.al.10}, we also derive the exact term linear in the mass ratio 
in $g^\text{eff}_{rr}$. In Sec.~\ref{sec:EOB-SF-PN} we use the binding energy computed through 6PN 
order in Ref.~\cite{Le.al2.11}, together with several constraints among the EOB metric coefficients 
derived in Refs.~\cite{Da.10,Ba.al.10}, to compute $g^\text{eff}_{tt}$ and $g^\text{eff}_{rr}$ through 6PN and 
5PN orders, respectively, at linear order in the mass ratio. Finally, in Sec.~\ref{sec:disc} we discuss 
the main results of the paper and comment on future work. The 
structure of the EOB Hamiltonian used in the rest of this paper is detailed in an Appendix.

\section{Self-force contributions to the EOB potentials}
\label{sec:EOB-SF-gen}

Within the EOB framework, the real Hamiltonian $H_{\rm real}$ encoding the orbital dynamics of 
two non-spinning compact objects is mapped to an {\it effective} Hamiltonian $H_{\rm eff}$ 
describing a test particle of mass $\mu = m_1 m_2/(m_1+m_2)$ moving in a \textit{deformed} 
Schwarzschild metric of mass $M = m_1 + m_2$. The deformation is regulated by the binary's 
symmetric mass ratio $\nu=\mu/M$, and disappears in the test-particle limit $\nu \rightarrow 0$. 
The EOB effective metric reads \cite{BuDa.99}
\begin{equation}\label{eq:EOBmetric}
	\ud s^2_\mathrm{eff} = - A(r) \, \ud t^2 + B(r) \, \ud r^2 + r^2 \ud \Omega^2 \, ,
\end{equation}
where the potentials $A \equiv - g^\text{eff}_{tt}$ and $B \equiv g^\text{eff}_{rr}$ are 
known through 3PN order \cite{BuDa.99,Da.al3.00}. We find it more convenient to work with the potential $\bar{D} \equiv (A B)^{-1}$, so that
\begin{subequations}\label{A_D_PN}
	\begin{align}
		A(u) &= 1 - 2u  + 2 \nu\,u^3 + \left (\frac{94}{3} - \frac{41}{32} \pi^2\right) \nu\,u^4 + \mathcal{O}(u^5) \, , \label{A_PN} \\
		\bar{D}(u) &= 1 + 6\nu\,u^2 + (52 - 6 \nu)\,\nu\,u^3 + \mathcal{O}(u^4) \, , \label{D_PN}
	\end{align}
\end{subequations}
where $u\equiv M/r$ denotes the inverse Schwarzschild-like EOB radial coordinate. In the test-particle limit $\nu \rightarrow 0$, we recover (by construction) the Schwarzschild results $A(u) = 1 - 2u$ and $\bar{D}(u) = 1$. Through 3PN order, the effective Hamiltonian is given by \cite{BuDa.99,Da.al3.00}
\begin{align}\label{Heff}
	H^2_{\rm eff}(\bm{r},\bm{p}) &= \mu^2 A(r) \, \bigl[ 1 + \hat{\bm{p}}^2 + \left( B(r)^{-1} - 1 \right) (\bm{n}\cdot\hat{\bm{p}})^2 \nonumber \\ &+ 2(4-3\nu) \, \nu \, u^2 \, (\bm{n}\cdot \hat{\bm{p}})^4 \bigr] \, ,
\end{align}
where we introduced the reduced conjugate momentum $\hat{\bm{p}} = \bm{p}/\mu$ and the unit vector $\bm{n} = \bm{r} / r$. Finally, the so-called improved real (or EOB) Hamiltonian reads \cite{BuDa.99}
\begin{equation}\label{Hreal}
	H_{\rm EOB} \equiv H^{\rm improved}_\mathrm{real} = M\,\sqrt{1+2\nu\left(\frac{H_{\rm eff}}{\mu}-1\right)}\,.
\end{equation}

When extending the EOB Hamiltonian \eqref{Hreal} to higher PN orders, 
one needs to modify the effective dynamics with terms depending on higher-order powers 
of the momentum $\hat{\bm{p}}$ \cite{Da.al3.00}, thus resulting in sextic and higher powers of $\hat{\boldsymbol{p}}$ inside the square brackets 
of the effective Hamiltonian (\ref{Heff}). However, as shown in Ref.~\cite{Da.al3.00} and discussed in the
Appendix, the mapping between the real and the effective descriptions can always 
be arranged in such a way that 
these ``non-geodesic'' terms are proportional to the 
radial momentum $\hat{p}_r \equiv \bm{n}\cdot \hat{\bm{p}}$, thus vanishing for circular orbits, at all PN orders. 
Moreover, we will show in the Appendix that this holds true even if the mapping  \eqref{Hreal} between the real and 
the effective Hamiltonians is assumed to be valid at \textit{all} PN orders. This 
standpoint was also adopted in Ref.~\cite{Da.10}, and it is the one
that we embrace in this paper. In other words, our results are valid for the class of EOB models
that adopt the mapping \eqref{Hreal} at all PN orders, and for which the non-geodesic higher-order momentum
terms are proportional to $\hat{p}_r$, thus vanishing for circular orbits.

We notice that the 3PN-accurate expression \eqref{A_PN} of the EOB potential $A(r)$ does not contain terms proportional to $\nu^2$ and $\nu^3$, despite the fact the PN binding energy does contain such terms. By contrast, at 3PN order, the EOB ``radial'' potential $B(r)$ --- and hence the inverse product $\bar{D} = (AB)^{-1}$ [\textit{cf.} Eq.~(\ref{D_PN})] --- contains a term proportional to $\nu^2$. The GSF results will allow us to control the exact contributions proportional to the binary's mass ratio $q \equiv m_1/m_2 =\nu + {\cal O}(\nu^2)$,\footnote{Without any loss of generality, we assume $m_1 \leqslant m_2$.} thus {\it only} the terms linear in $\nu$ in $A(r)$ and $\bar{D}(r)$.

In the next two Subsections, we will derive the GSF contributions to the EOB potentials, 
namely the functions $A_{\rm SF}(u)$ and $\bar{D}_{\rm SF}(u)$ such that\footnote{In the notations of Refs.~\cite{Da.10,Ba.al.10}, we have $A_{\rm SF} = a$ and $\bar{D}_{\rm SF} = \bar{d}$.}
\begin{subequations}
	\begin{align}
		A(u) &= 1 - 2u + \nu \, A_{\rm SF}(u) + {\cal O}(\nu^2) \, , \label{potentialA} \\
		\bar{D}(u) &= 1 + \nu \, \bar{D}_{\rm SF}(u) + {\cal O}(\nu^2) \, . \label{potentialD}
	\end{align}
\end{subequations}

\subsection{Self-force contribution to the EOB effective metric potential $A$}

We will restrict the discussion to \textit{circular} orbits, computing 
first the EOB energy $E_{\rm EOB}$ for such orbits. 
Now, the angular momentum $L \equiv p_\phi$ can be determined as a function of 
the inverse separation $u$ by solving the equation
\begin{equation}
\label{dotpr}
\dot{p}_r = - \frac{\partial H_{\rm EOB}}{\partial r}(r,p_r=0,p_\phi)=0\,,
\end{equation}
which is valid for circular orbits only.
From the expressions \eqref{Heff} and \eqref{Hreal} of the effective and EOB Hamiltonians, this gives
\begin{equation}\label{L}
	\frac{L^2(u)}{\mu^2 M^2} = - \frac{A'(u)}{2u A(u) + u^2 A'(u)} \, ,
\end{equation}
where we denote $A' \equiv \ud A / \ud u$. Replacing this result in the expressions for the Hamiltonians, we obtain the circular-orbit EOB energy as
\begin{subequations}\label{eEOBu}
	\begin{align}
		E_{\rm EOB}(u) &= M \,\sqrt{1+2\nu\left(\frac{E_{\rm eff}(u)}{\mu}-1\right)} \,,\\
		E_{\rm eff}(u) &= \mu\,\sqrt{\frac{2A^2(u)}{2A(u)+u A^{\prime}(u)}}\,.
	\end{align}
\end{subequations}
We also introduce the dimensionless coordinate-invariant quantity 
$x \equiv (M\Omega)^{2/3}$, where $\Omega$ is the constant circular-orbit frequency, given by 
\begin{equation}
\label{Omega}
\Omega = \frac{\partial H_{\rm EOB}}{\partial p_\phi}(r,p_r=0,p_\phi)\,,
\end{equation}
as well as $r_\Omega \equiv M / x$, a convenient invariant measure of the orbital separation.

Since we are interested in improving the EOB Hamiltonian using GSF results, we will work at linear order in the symmetric mass ratio $\nu$, thus neglecting terms ${\cal O}(\nu^2)$ or higher. Inverting Eq.~(\ref{Omega}) and making use of the expression \eqref{L} for the angular momentum
 yields \cite{Da.10}
\begin{equation}
\label{uversusx}
u = x + \nu \left[ \frac{x}{6} \, A_\text{SF}'(x) + \frac{2}{3} \, x \left( \frac{1-2x}{\sqrt{1-3x}} -1 \right) \right] + {\cal O}(\nu^2) \, .
\end{equation}
Next, we compute the (specific) EOB binding energy $\hat{E}_{\rm EOB} \equiv (E_{\rm EOB} - M) / \mu$ 
at linear order in $\nu$, for circular orbits. From Eqs.~\eqref{eEOBu}, in which we 
substitute the coordinate-dependent variable $u$ in favor of the invariant variable $x$ using \eqref{uversusx}, we obtain
\begin{align}\label{eEOBx}
	&\hat{E}_{\rm EOB}(x) = \frac{1-2x}{\sqrt{1-3x}} - 1 + \nu \, \bigg\{ \frac{1-4x}{(1-3x)^{3/2}} \frac{A_{\rm SF}(x)}{2} \nonumber \\ &- \frac{x}{\sqrt{1-3x}} \frac{A'_{\rm SF}(x)}{3} - \biggl( \frac{1-2x}{\sqrt{1-3x}} -1 \biggr) \, \times \nonumber \\ & \biggl[ \frac{x}{3} \frac{1-6x}{(1-3x)^{3/2}} + \frac{1}{2} \biggl( \frac{1-2x}{\sqrt{1-3x}} - 1 \biggr) \biggr] \biggr\} + {\cal O}(\nu^2)\,.
\end{align}

Recently, Ref.~\cite{Le.al3.11} used the first law of binary point-particle mechanics \cite{Le.al2.11}, together with GSF results for the redshift observable $z$ \cite{De.08,Sa.al.08,Sh.al.11,Bl.al.10,Bl.al2.10}, to compute the binding energy $E$ and total angular momentum $L$ of a circular-orbit non-spinning compact binary, through next-to-leading order in the symmetric mass ratio $\nu$ (at all PN orders). In particular, the specific binding energy $\hat{E} \equiv E / \mu$ reads \cite{Le.al3.11}
\begin{align}\label{eSFx}
	\hat{E}(x) &= \frac{1 -2x}{\sqrt{1-3x}} - 1 + \nu \, \biggl[ \frac{1}{2}\,z_{\rm SF}(x) - \frac{x}{3}\,z'_{\rm SF}(x) \nonumber \\ &+ \sqrt{1-3x} - 1 + \frac{x}{6} \, \frac{7-24x}{(1-3x)^{3/2}} \biggr] + {\cal O}(\nu^2) \, ,
\end{align}
where $z_\text{SF}(x)$ is the self-force contribution to the redshift $z$ of the lightest point mass, which is known numerically, with high accuracy, for circular orbits down to $r_\Omega = 5M$ (see below for more details).

By construction of the EOB model, the EOB binding energy $\hat{E}_{\rm EOB}$ must 
coincide with the binding  energy $\hat{E}$ of the real binary system when expanded in 
powers of $\nu$. Equating Eqs.~\eqref{eEOBx} and \eqref{eSFx} yields the following linear 
first-order ordinary differential equation for $A_{\rm SF}(x)$:
\begin{align}
	& 2x\, A'_{\rm SF}(x) - 3 \frac{1-4x}{1-3x} A_{\rm SF}(x) = x \,\frac{1-6x}{1-3x} + \sqrt{1-3x} \, \times \nonumber \\ &\left[ 2x \, z'_{\rm SF}(x) - 3 z_{\rm SF}(x) + x\, \frac{1-5x+12x^2}{(1-3x)^2} \right] .
\end{align}
Interestingly, this differential equation can be solved analytically in terms of 
$z_{\rm SF}(x)$. The solution is particularly simple, and explicitly 
reads\footnote{The homogeneous solution must vanish because the PN expansion of $A_{\rm SF}(x)$ cannot involve half-integer powers of $x$.}
\beq\label{ASF}
	A_{\rm SF}(x) = \sqrt{1-3x} \,\, z_{\rm SF}(x) - x \left( 1 + \frac{1-4x}{\sqrt{1-3x}} \right) .
\eeq
This is one of the most important results of this paper: we have succeeded in relating the 
known GSF contribution $z_{\rm SF}(x)$ in the redshift observable to the EOB dynamics for circular orbits. 
As a result, we can now compute the EOB radial potential $A(r)$ given in Eq.~(\ref{potentialA}) including {\it all} PN corrections, 
at linear order in $\nu$. As a consistency check, the PN expansion of $z_{\rm SF}(x)$, as given by Eq.~(4.16) and Table~I of Ref.~\cite{Le.al2.11}, 
can be used with Eq.~\eqref{ASF} to recover the 3PN expansion \eqref{A_PN} of $A_{\rm SF}(x)$.

We stress that the EOB model with this form of the potential $A(r)$ automatically reproduces
the shift of the ISCO frequency under the effect of the conservative part of the GSF,
as calculated by Barack and Sago in Refs.~\cite{BaSa.09,BaSa.10}. This is because the notion of ISCO coincides with that of minimum energy
circular orbit (MECO) for any orbital dynamics that can be derived from a Hamiltonian \cite{Bu.al.03}, and the MECO as computed
from Eq.~\eqref{eSFx} returns the correct ISCO frequency shift \cite{Le.al3.11}.
This is a considerable improvement over earlier versions of the EOB potential $A(r)$, which were 
resorting to free parameters regulating unknown high-order PN effects \cite{Da.10,BaBu.10,Pa.al.11} 
in order to reproduce the ISCO frequency shift due to the conservative GSF.

Furthermore, if we use our newly derived potential $A(r)$ together with Eqs.~\eqref{L} and \eqref{uversusx} to compute 
the dimensionless angular momentum $\hat{L} \equiv L / (\mu M)$ for circular orbits, expressed in terms of the coordinate-invariant quantity $x$,
we recover the result obtained in Ref.~\cite{Le.al3.11}, namely
\begin{align}
\hat{L}(x) &= \frac{1}{\sqrt{x(1-3x)}} + \nu \, \biggl[ - \frac{1}{3\sqrt{x}} \, z'_{\rm SF}(x) \nonumber \\ &\qquad\qquad + \frac{1}{6\sqrt{x}} \frac{4-15x}{(1-3x)^{3/2}} \biggr] + {\cal O}(\nu^2) \, ,
\end{align}
which holds at all PN orders and at linear order in $\nu$. This comes at no surprise because the binding energy and total angular momentum 
for circular orbits satisfy the exact relation
$\partial \hat{E}/\partial \hat{L} = M \Omega$ in the EOB model. Therefore, once the energy 
coincides with the exact expression established in Ref.~\cite{Le.al3.11}, 
so must the angular momentum.

The GSF contribution $z_{\rm SF}(x)$ to the redshift observable has been calculated numerically in Refs.~\cite{De.08,Sa.al.08,Sh.al.11,Bl.al.10,Bl.al2.10} for a variety of orbital separations, in the range $5M \leqslant r_\Omega \leqslant 500M$. In the companion paper \cite{Le.al3.11}, we established that this numerical data can conveniently be represented with an accuracy better than a part in $10^{5}$, using the compact analytical formula
\beq\label{zSF}
z_{\rm SF}(x) = 2 x \, \frac{1 + a_1 x + a_2 x^2}{1 + a_3 x + a_4 x^2 + a_5 x^3} \, ,
\eeq
with the coefficients $a_1 = -2.18522$, $a_2 = 1.05185$, $a_3 = -2.43395$, $a_4 = 0.400665$, and $a_5 = -5.9991$. 
This fitting formula accounts for the leading-order (1PN) behavior $z_{\rm SF}(x) = 2 x + {\cal O}(x^2)$ when $x \to 0$ \cite{Le.al2.11}. 
Since current GSF data for $z_{\rm SF}(x)$ is limited to orbital separations $r_\Omega \geqslant 5M$, 
the GSF-accurate EOB potential $A(u)$ given by Eqs.~\eqref{potentialA} and \eqref{ASF} is (for now) only known in the range $0 \leqslant x \leqslant 1/5$. 

\subsection{Self-force contribution to the EOB effective metric potential $\bar{D}$}

The non-circular conservative dynamics of spinless binaries is regulated by
the radial frequency $\Omega_r$ and by the averaged angular
frequency $\Omega_\phi$, respectively defined by
\begin{subequations}\label{Omegas}
	\begin{align}
		\Omega_r &\equiv \frac{2 \pi}{P} \, , \\
		\Omega_\phi &\equiv \frac{1}{P} \int_0^P \! \dot{\phi}(t) \, \ud t = K \, \Omega_r \, ,
	\end{align}
\end{subequations}
where $P$ is the radial period, namely the time interval between two
successive periastron passages, $\dot{\phi} \equiv \ud \phi / \ud t$
is the time derivative of the orbital phase $\phi(t)$, and $\Delta
\Phi / (2 \pi) = K-1$ is the fractional advance of the periastron per
radial period. In the circular-orbit limit, by definition the radial frequency 
vanishes at the ISCO; hence the periastron advance 
$K = \Omega_\phi / \Omega_r$ blows up there. For this reason,  Refs.~\cite{Da.10,Ba.al.10} found 
it convenient to work with the quantity $W \equiv {1}/{K^2}$, which is regular at the ISCO. Reference \cite{Ba.al.10} calculated numerically the GSF contribution to $W$, \textit{i.e.} the function $\rho_{\rm SF}(x)$ such that
\begin{equation}\label{KSF}
	W(x) = 1 - 6x +\nu\,\rho_{\rm SF}(x) + {\cal O}(\nu^2) \, .
\end{equation}
The authors performed several fits of the
GSF data for $\rho_{\rm SF}(x)$ (in the range $6M \leqslant r_\Omega
\leqslant 80M$). In particular, they found that this data can be
accurately reproduced at the $2.4 \times 10^{-3}$ level by means of
the compact analytic formula
\begin{equation}\label{rhoSF}
\rho_{\rm SF}(x) = 14 x^2 \frac{1 + b_1\, x}{1 + b_2\, x + b_3\, x^2} \, ,
\end{equation}
with $b_1 = 13.3687$, $b_2 = 4.60958$, and $b_3 = -9.47696$. (Using a denser data set in a more limited frequency range, Ref.~\cite{Le.al.11} later found that the values $b_1 = 12.9906$, $b_2 = 4.57724$, and $b_3 = -10.3124$ yield a fit accurate at the $10^{-5}$ level.) As with Eq.~\eqref{zSF}, the fitting formula \eqref{rhoSF} accounts for the leading-order 
(2PN) behavior of $\rho_{\rm SF}(x)$ when $x \to 0$ \cite{Ba.al.10}.

Reference \cite{Da.10} recently studied the dynamics of slightly
eccentric orbits within the EOB formalism, and found that the GSF
correction $\rho_{\rm SF}(x)$ to the periastron advance is related to
the EOB potentials $A_{\rm SF}(x)$ and $\bar{D}_{\rm SF}(x)$
by\footnote{This result requires that the quartic power in the radial momentum $p_r$ inside the square brackets of Eq.~(\ref{Heff}) be neglected. This is correct in the limit of small eccentricity $e \ll 1$, for which $p_r \sim e$; see Ref.~\cite{Da.10} for more details.}
\begin{align}
\label{rhoAD}
\rho_{\rm SF}(x) &= 
4x \left( 1 - \frac{1-2x}{\sqrt{1-3x}} \right) + A_{\rm SF}(x) 
+ x \, A'_{\rm SF}(x) \nonumber \\ &+ \frac{x}{2} \, (1-2x) \, A''_{\rm SF}(x) + (1-6x) \, \bar{D}_{\rm SF}(x) \, .
\end{align}
Solving the above equation for the unknown $\bar{D}_{\rm SF}(x)$, we immediately obtain
\begin{align}\label{DSF}
&\bar{D}_{\rm SF}(x) = \frac{1}{1-6x} \, \biggl[ \rho_{\rm SF}(x) 
+ 4x \left( \frac{1-2x}{\sqrt{1-3x}} - 1 \right) \nonumber\\&\quad- A_{\rm SF}(x) 
- x \, A'_{\rm SF}(x) - \frac{x}{2} \, (1-2x) \, A''_{\rm SF}(x) 
\biggr] \, ,
\end{align}
where $A_{\rm SF}(x)$ is given explicitly in terms of $z_\text{SF}(x)$ by Eq.~(\ref{ASF}) above. Equation (\ref{DSF}) is another important result of this paper: the EOB potential $\bar{D}(r)$ governing the \textit{non-radial} motion, as given by Eq.~(\ref{potentialD}), is now known exactly at linear order in $\nu$, through the known GSF contributions $z_\text{SF}(x)$ and $\rho_\text{SF}(x)$ to the redshift observable and periastron advance.

Note that the apparent pole at the Schwarzschild ISCO
($x = 1/6$) in Eq.~\eqref{DSF} must be canceled out by a factor $(1-6x)$ in the numerator, because the potential $\bar{D}(r)$ is
perfectly regular at the ISCO. This can be verified using the fit
\eqref{zSF} for $z_{\rm SF}(x)$, which is accurate to within $10^{-5}$,
together with the fit \eqref{rhoSF} for $\rho_{\rm SF}(x)$, which is accurate to within $2.4\times 10^{-3}$ using 
the coefficients $b_1$, $b_2$ and $b_3$ from Ref.~\cite{Ba.al.10}, and to within $10^{-5}$, although in a more
limited frequency range, using the coefficients $b_1$, $b_2$ and $b_3$ from Ref.~\cite{Le.al.11}.
Using these fits, the behavior of $\bar{D}_{\rm SF}(x)$
near the ISCO is of the form $\bar{D}_{\rm SF}(x)= \epsilon/(x-1/6)+{\cal O}(1)$, where the 
dimensionless parameter $\epsilon$ is about $4\times10^{-5}$ with 
the coefficients $b_1$, $b_2$ and $b_3$ from Ref.~\cite{Ba.al.10}, and $3.4\times10^{-5}$ 
with the coefficients $b_1$, $b_2$ and $b_3$ from Ref.~\cite{Le.al.11}.
One can therefore argue that the value of $\epsilon$ is comparable to the accuracy of the fits
and thus compatible with zero. 

Finally, when using the known PN expansions of the GSF contributions $\rho_{\rm SF}(x)$ and $z_{\rm SF}(x)$ to the redshift observable and periastron advance (see \textit{e.g.} Eqs.~(5.32) and (5.39) of Ref.~\cite{Da.10}, and Eq.~(4.16) and Table~I of Ref.~\cite{Le.al2.11}), we recover, as expected, the 3PN expansion \eqref{D_PN} of $\bar{D}_{\rm SF}(x)$.

\section{High-order post-Newtonian terms in the EOB potentials}
\label{sec:EOB-SF-PN}

Recently, the authors of Ref.~\cite{Le.al2.11} derived a first
  law of mechanics for non-spinning compact objects modelled as point particles, and moving along
  exactly circular orbits. By making use of previous GSF results for
  the redshift observable \cite{De.08,Bl.al.10,Bl.al2.10}, they could determine the numerical values of some
  previously unknown coefficients entering the expression of the
  circular-orbit binding energy $E$. Through 6PN order, the (specific)
  binding energy explicitly reads 
\begin{widetext}
	\begin{align}\label{EPN}
		\hat{E}(x) = - \frac{x}{2} & \, \left\{ 1 + \left( -\frac{3}{4} - \frac{\nu}{12} \right) x + \left( - \frac{27}{8} + \frac{19}{8} \nu - \frac{\nu^2}{24} \right) x^2 + \left( - \frac{675}{64} + \left[ \frac{34445}{576} - \frac{205}{96} \pi^2 \right] \nu \right. \right. \nonumber \\ &\quad\; \left. \left. - \, \frac{155}{96} \nu^2 - \frac{35}{5184} \nu^3 \right) x^3 + \left( - \frac{3969}{128} + \nu \,e_4(\nu) + \frac{448}{15}\nu\,\ln{x} \right) x^4
+ \left( - \frac{45927}{512} + \nu \,e_5(\nu) \right. \right. \nonumber \\ &\quad\; \left. \left. + \left[ - \frac{4988}{35} - \frac{1904}{15} \nu \right] \nu \, \ln x \right) x^5 + \left( - \frac{264627}{1024} + \nu \,e_6(\nu) + \nu \, e^{\rm ln}_6(\nu) \, \ln{x} \right) x^6 + o(x^6) \right\} ,
	\end{align}
\end{widetext}
where the numerical values of the 4PN, 5PN and 6PN coefficients $e_4(0)$, $e_5(0)$, $e_6(0)$, and $e^{\rm ln}_6(0)$ 
were found to be~\cite{Le.al2.11}
\begin{subequations}\label{PNcoeff}
	\begin{align}
		e_4(0) &= +153.8803(1)\,,\\
		e_5(0) &= -55.13(3)\,,\\
		e_6(0) &= +588(7)\,, \\
		e^{\rm ln}_6(0) &= -1144(2)\,.
	\end{align}
\end{subequations}
(The uncertainty in the last digit is indicated in parenthesis.) Note
that the leading-order 4PN and next-to-leading order 5PN logarithmic
contributions to the binding energy are known analytically \cite{Bl.al2.10}. The value
of the post-GSF coefficient (term $\propto \nu^2$) in the 5PN
logarithmic contribution is that corresponding to the ``physical
problem'' in the language of Ref.~\cite{Le.al2.11}, \textit{i.e.}
when the helical symmetry is \textit{not} imposed (see
\cite{Le.al2.11} for more details).

In the next two Subsections, we will use Eqs.~\eqref{EPN} and
\eqref{PNcoeff}, together with the results of
Ref.~\cite{Ba.al.10}, to compute the 4PN, 5PN and 6PN
coefficients in $A(u)$, as well as the 4PN and 5PN coefficients in
$\bar{D}(u)$, at linear order in the symmetric mass ratio $\nu$.

\subsection{High-order post-Newtonian terms in the EOB metric potential $A$}
\label{sec:potentialA}

Similarly to what was done in Sec.~\ref{sec:EOB-SF-gen}, we restrict here to the class of EOB models
that adopt the mapping \eqref{Hreal} at all PN orders, and for which the non-geodesic higher-order momentum
terms are proportional to the radial momentum. As a consequence, the 4PN, 5PN and 6PN corrections determined in Eqs.~\eqref{EPN} and \eqref{PNcoeff} enter the effective Hamiltonian only through the radial potential 
$A(r)$, \textit{i.e.}, through the coefficients $a_5(\nu)$, $a_5^{\ln}(\nu)$, $a_6(\nu)$, $a_6^{\ln}(\nu)$, 
$a_7(\nu)$ and $a_7^{\ln}(\nu)$ in the PN expansion
\begin{align}\label{A6PN}
	A(u) = 1 &- 2u + 2\nu\,u^3 + \left ( \frac{94}{3} - \frac{41 \pi^2}{32} \right) \nu\,u^4 \nonumber \\
		&+ \nu \left [a_5(\nu) + a_5^{\ln}(\nu)\,\ln{u} \right] u^5 \nonumber \\
		&+ \nu \left [a_6(\nu) + a_6^{\ln}(\nu)\,\ln u \right] u^6\nonumber \\
		&+ \nu \left [a_7(\nu) + a_7^{\ln}(\nu)\,\ln u \right] u^7 + o(u^7) \, .
\end{align}
A comment regarding the general structure of the PN expansion of
$A(u)$ is in order. It was argued in Ref.~\cite{Bl.al2.10} that
terms involving powers of logarithms should not occur in the
\textit{conservative} part of the dynamics of a compact binary system
before the very high 7PN order. Within the class of EOB Hamiltonians
that we consider in this work, the potential $A$ is directly related
to the conserved binding energy $E$ of the real binary system. Hence
we expect no term of the type $(\ln{u})^p$, with $p \geqslant 2$, in
the 6PN-accurate expansion \eqref{A6PN} of $A(u)$.

Now, in order to derive the $\nu \to 0$ limits of the coefficients
$a_i(\nu)$ and $a_i^{\ln}(\nu)$ (with $i=5,6,7$), we first determine
the angular momentum $L = p_\phi$ as a function of the circular-orbit
frequency $\Omega$ by solving Eqs.~(\ref{dotpr}) and (\ref{Omega}).
Second, we insert $L=L(\Omega)$ in Eqs.~(\ref{Heff}) and
(\ref{Hreal}), set $p_r=0$, and expand Eq.~(\ref{Hreal}) in PN
orders. Third, after subtracting the rest-mass contribution, we equate
the result to Eq.~(\ref{EPN}), and finally obtain
\begin{subequations}\label{coeffASF}
	\begin{align}
		a_5(0) &= -\frac{275\,139}{4\,480} + \frac{3}{7}\,e_4(0) + \frac{123}{64}\pi^2\,,\\
		a_5^{\ln}(0) &= \frac{64}{5}\,,\label{a5ln}\\
		a_6(0) &= -\frac{2\,772\,125}{96\,768} - \frac{9}{14}\,e_4(0) \nonumber \\ &\quad + \frac{1}{3}\,e_5(0) + \frac{369}{256}\pi^2\,,\\
		a_6^{\ln}(0) &= - \frac{7\,004}{105}\,,\label{a6ln}\\
		a_7(0) &= -\frac{23\,821\,223}{322\,560} -\frac{27}{56}\,e_4(0) - \frac{1}{2}e_5(0) \nonumber \\ &\quad + \frac{3}{11}\,e_6(0) - \frac{6}{121}\,e_6^{\ln}(0) + \frac{1\,107}{512}\pi^2\,,\\
		a_7^{\ln}(0) &= \frac{398}{7} + \frac{3}{11}\,e_6^{\ln}(0)\,.
	\end{align}
\end{subequations}
Note that the values of the 4PN and 5PN log-coefficients $a_5^{\ln}(0)$ and $a_6^{\ln}(0)$ are known analytically.\footnote{Actually these logarithmic terms are known for \textit{all} mass ratios: $a_5^{\ln}(\nu) = \frac{64}{5}$ and $a_6^{\ln}(\nu) = - \frac{7\,004}{105} - \frac{144}{5} \, \nu$.} Using the numerical results \eqref{PNcoeff} for the binding energy, we find for the other 4PN, 5PN and 6PN unknown coefficients:
\begin{subequations}\label{APNcoeff}
	\begin{align}
		a_5(0) &= +23.50190(5)\,,\label{a5}\\
		a_6(0) &= -131.72(1)\,,\label{a6}\\
		a_7(0) &= +118(2)\,,\\
		a_7^{\ln}(0) &= -255.0(5)\,.
	\end{align}
\end{subequations}
As a consistency check, we verified that the same values for these coefficients also follow from
expanding in PN orders the potential $A$ given by Eqs.~\eqref{potentialA} and \eqref{ASF}, using the 6PN-accurate fit to 
$z_{\rm SF}(x)$ given in Eq.~(4.16) and Table II of Ref.~\cite{Le.al2.11}.

Finally, we note that the values of the coefficients $a_5(0)$ and
  $a_6(0)$, as determined in Ref.~\cite{Da.10} [see Eq.~(4.46)
  there] by combining the EOB/NR comparison of Ref.~\cite{DaNa.09}
  with the constraint coming from the GSF calculation of the ISCO
  frequency shift, are in poor agreement with the exact results
  \eqref{a5} and \eqref{a6}; in particular the signs of both coefficients
  are wrongly predicted. Even without taking into account the
  constraint from the GSF, the correct values \eqref{a5} and \eqref{a6}
  lie far outside the (banana-shaped) region of the $a_5(0),a_6(0)$
  plane favored by a comparison of the EOB prediction with the NR
  waveform of an equal-mass binary black-hole simulation
  \cite{DaNa.09,DaNa.11}. This disagreement does not surprise us. Indeed, several causes can affect the correct values of the coefficients $a_5(0)$ and $a_6(0)$ when they are extracted from a EOB/NR calibration, such as radiation-reaction effects and all higher-order PN contributions in $A(r)$, which become important during the plunge and close to merger, but are neglected in those calibrations.

\begin{figure*}
\begin{center}
  \includegraphics[width=0.47\linewidth]{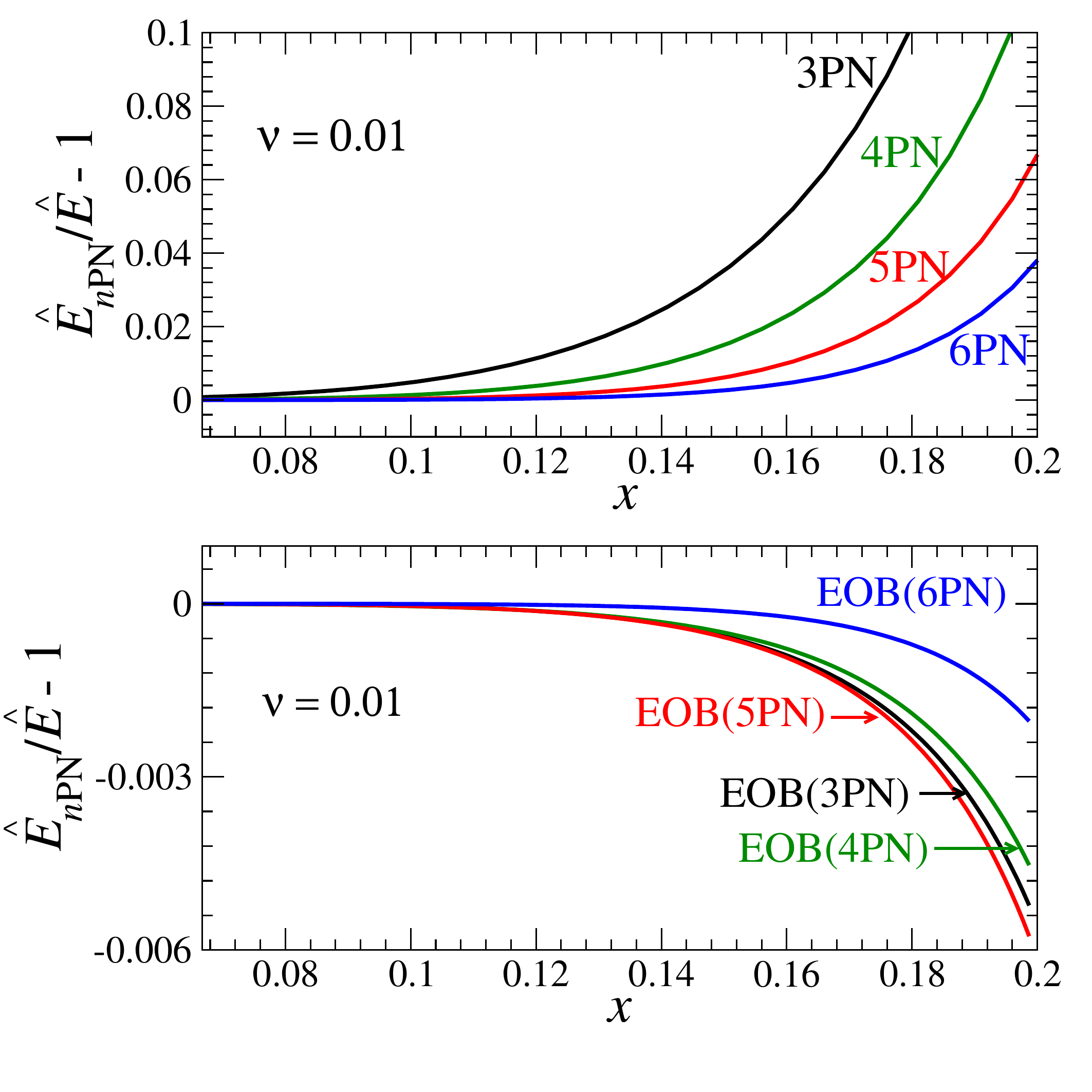} \quad
  \includegraphics[width=0.47\linewidth]{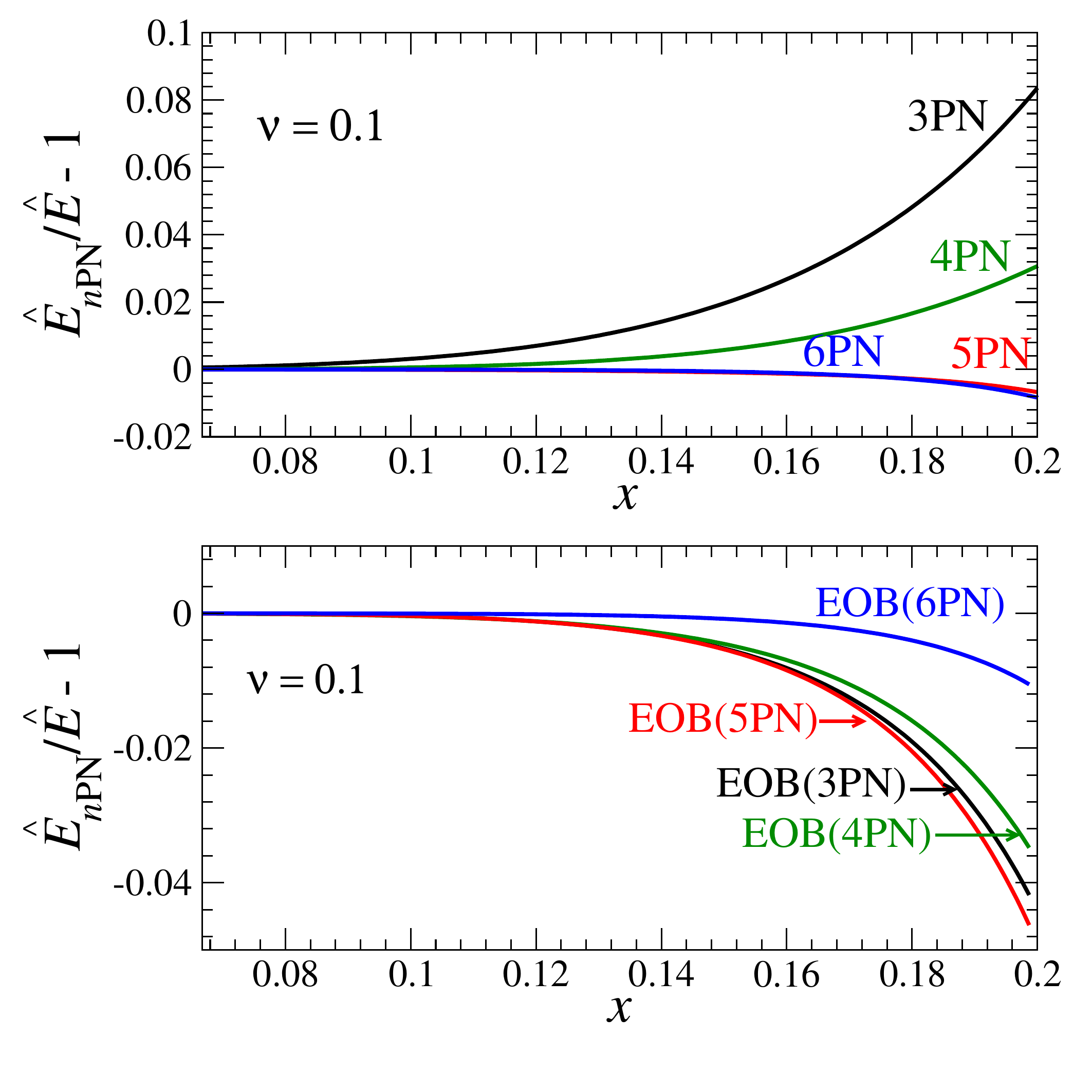} \vspace{-0.7cm}
\end{center}
  \caption{The fractional difference between the GSF-accurate 
(specific) binding energy and the (specific) binding energy calculated at linear order in $\nu$ using 
standard PN approximants (upper plot), and EOB approximants (lower plot), for 
$\nu = 0.01$ (left panel) and $\nu = 0.1$ (right panel).}
\label{fig:E001-01}
\end{figure*}
\begin{figure}
\includegraphics[width=\linewidth]{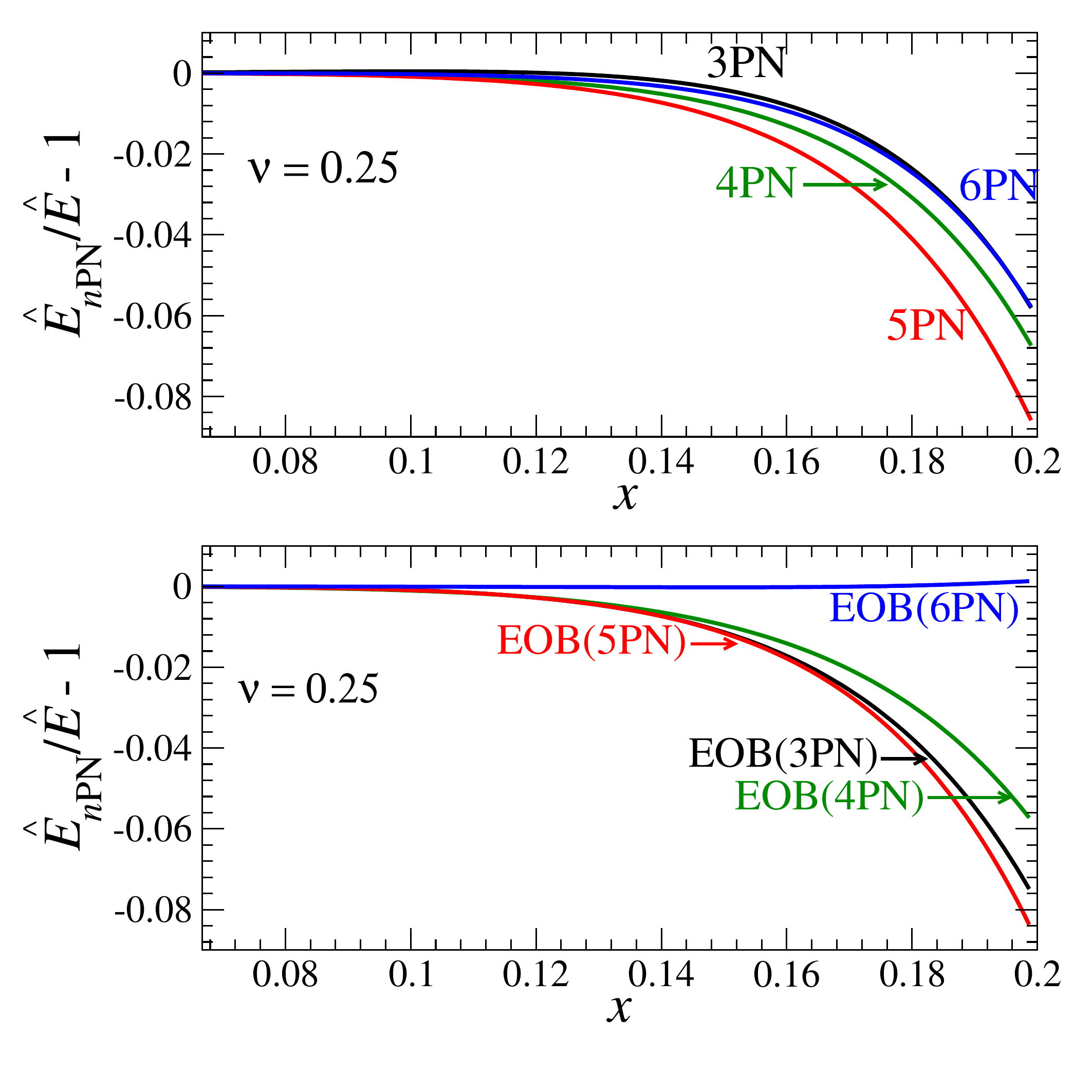} \vspace{-0.7cm}
\caption{Same as in Fig.~\ref{fig:E001-01}, for an equal-mass binary ($\nu=0.25$).}
\label{fig:E025}
\end{figure}

\subsection{High-order post-Newtonian terms in the EOB metric potential $\bar{D}$}
\label{sec:potentialD}

We will now show that by combining the results of this paper with those of Ref.~\cite{Ba.al.10}, 
we can also determine the exact or approximate numerical values of the non-logarithmic and logarithmic coefficients $\bar{d}_4(\nu)$, $\bar{d}_5(\nu)$, $\bar{d}_4^{\ln}(\nu)$ and $\bar{d}_5^{\ln}(\nu)$ 
appearing at 4PN and 5PN order in the potential $\bar{D}$:
\begin{align}\label{ID5PN}
	\bar{D}(u) = 1 &+ 6\nu\,u^2 + (52 - 6 \nu)\,\nu\,u^3 \nonumber \\
			&+ \nu \left [\bar{d}_4(\nu) + \bar{d}_4^{\ln}(\nu)\,\ln{u} \right] u^4 \nonumber \\
			&+ \nu \left [\bar{d}_5(\nu) + \bar{d}_5^{\ln}(\nu)\,\ln{u} \right] u^5 + o(u^5)\,.
\end{align}
The authors of Ref.~\cite{Ba.al.10} extracted numerically some unknown high-order PN 
terms in the function $\rho_{\rm SF}(x)$, and used the relation \eqref{rhoAD} between $\rho_{\rm SF}$, $A_{\rm SF}$, and $\bar{D}_{\rm SF}$ 
to put constraints on some unknown high-order PN coefficients appearing in the EOB potentials $A(u)$ and $\bar{D}(u)$. In particular, Ref.~\cite{Ba.al.10} derived the following constraints on the $\nu \to 0$ limit of the PN coefficients $\bar{d}_4(\nu)$ and $\bar{d}_5(\nu)$:
\begin{subequations}
	\begin{align}
		& 10\,a_5(0) + \bar{d}_4(0) + \frac{9}{2}\,a_5^{\ln}(0) = 518.6^{+7}_{-4}\,,\\
		& 14\,a_5(0) +6\,\bar{d}_4(0) -15\,a_6(0) - \bar{d}_5(0) \nonumber \\
		& + 8\,a_5^{\ln}(0)-\frac{11}{2}\,a_6^{\ln}(0) = 4779^{-400}_{+1200}\,,
\end{align}
\end{subequations}
as well as the following exact relations between the $\nu \to 0$ limit of the PN coefficients $\bar{d}_4^{\ln}(\nu)$ and $\bar{d}_5^{\ln}(\nu)$:
\begin{subequations}
	\begin{align}
		& 10\,a_5^{\ln}(0) + \bar{d}_4^{\ln}(0) = \frac{2\,512}{15}\,,\\
		& 14\,a_5^{\ln}(0) - 15\,a_6^{\ln}(0) + 6\,\bar{d}_4^{\ln}(0) - \bar{d}_5^{\ln}(0) = \frac{11\,336}{7}\,.
	\end{align}
\end{subequations}
Substituting Eqs.~\eqref{a5ln}, \eqref{a6ln}, and (\ref{APNcoeff}) in the above equations, we get
\begin{subequations}\label{coefsD}
	\begin{align}
		\bar{d}_4(0) &= +226.0^{+7}_{-4}\,,\\
		\bar{d}_4^{\ln}(0) &= +\frac{592}{15}\,,\\
		\bar{d}_5(0) &= - 649^{-1200}_{+400}\,,\\
		\bar{d}^{\ln}_5(0) &= - \frac{1420}{7}\,.
	\end{align}
\end{subequations}
As a consistency check, we verified that the values \eqref{coefsD} can be recovered by inserting
the PN fits to $z_{\rm SF}(x)$ and $\rho_{\rm SF}(x)$ given in Refs.~\cite{Le.al2.11,Ba.al.10} into the 
exact expression \eqref{potentialD} and \eqref{DSF} for $\bar{D}$.

In summary, building on the results of Refs.~\cite{Le.al2.11,Ba.al.10}, we 
have computed the 4PN, 5PN and 6PN terms in the EOB potential $A(u)$, as well as the 4PN and 5PN terms in the potential $\bar{D}(u)$,
at linear order in $\nu$.

\section{Discussion and conclusions}
\label{sec:disc}

The potential $A(u)$ given by Eqs.~\eqref{potentialA} and \eqref{ASF} ensures that the
EOB binding energy for circular orbits coincides, at linear order in $\nu$, with the exact result~(\ref{eSFx}).
In order to investigate the properties of the EOB resummation, as opposed to the standard
PN expansion, we will here compare this exact result for the binding energy to the PN predictions,
as well as to the EOB binding energy, as computed with the PN-expanded version of the potential $A(u)$
[\textit{cf.} Eq.~\eqref{A6PN}].

In Figs.~\ref{fig:E001-01} and \ref{fig:E025} we show the fractional difference 
between the GSF-accurate (specific) binding energy $\hat{E}(x)$ given by Eq.~(\ref{eSFx}), and either the 
EOB (specific) binding energy obtained from Eqs.~(\ref{eEOBu}) using the PN-expanded potential $A$ 
(at linear order in $\nu$), or the PN (specific) binding energy 
given in Eq.~(\ref{EPN}), including only the terms linear in $\nu$. The fractional difference is presented 
as a function of $x = (M \Omega)^{2/3}$ up to $x = 1/5$. We consider three representative symmetric mass ratios, namely $\nu = 0.01, 0.1$, and $0.25$, and several 
PN orders.\footnote{In order to express the EOB binding energy (\ref{eEOBu}) at $n$PN order as a function of $x$ rather than $u$, we 
insert Eq.~(\ref{L}) into the relation $x^{3/2} = M {\partial H_{\rm EOB}}/{\partial L}(r=1/u,p_r=0,L)$, 
and invert the latter to obtain $u$ as a PN expansion in $x$.}

From Figs.~\ref{fig:E001-01} and \ref{fig:E025} we observe that up to $x = 1/5$ the EOB-approximants 
are much closer than the PN-approximants to the exact GSF result for the 
small mass-ratio case $\nu = 0.01$, and are (roughly) comparable, with some differences 
depending on the PN order, to the exact GSF result for the comparable mass-ratio cases 
$\nu=0.1$ and $\nu=0.25$.

These results confirm the utility of resumming the PN dynamics around the test-mass
limit if the goal is to obtain a flexible model that can smoothly {\it bridge} between the 
test-mass and equal-mass limits. This flexibility exists in the 
EOB model not only at the level of the binding energy for circular orbits, but 
more importantly at the level of the Hamiltonian, thus for generic orbits and beyond the 
innermost stable circular orbit. It is a crucial feature that has allowed to build faithful inspiral-merger-ringdown 
templates that can span the entire binary's mass-ratio range \cite{Bu.al.07,Bu.al2.07,Pa.al.08,Bo.al.08,Bu.al.09,Pa.al.10,
DaNa.08,Da.al3.08,Da.al4.08,DaNa.09,Pa.al.11,DaNa.07,Na.al.07,BeNa.10,Be.al.11,Be.al2.11,Pa.al2.11,Yu.al.10,Yu.al.11,Ba.al2.12}.

We leave to future work a detailed study of how the EOB model,
augmented with the gravitational self-force results in $A(u)$ and
$\bar{D}(u)$, performs against numerical-relativity simulations of
comparable-mass black-hole binaries
\cite{Pa.al.11}. Comparisons using a 4PN, 5PN, or even 6PN-accurate EOB model can
already be pursued. However, in order to use the EOB potentials with
all PN terms linear in $\nu$, more self-force data is needed for the
redshift observable $z_{\rm SF}(x)$ beyond $x = 1/5$, and for the
periastron advance $\rho_{\rm SF}(x)$ beyond $x = 1/6$.

\begin{acknowledgments} 
All three authors acknowledge support from NSF Grant No. PHY-0903631. A.B. also acknowledges support 
from NASA Grant No. NNX09AI81G, and A.L.T. from the Maryland Center for Fundamental Physics.
\end{acknowledgments} 

\appendix

\section{On the generic structure of the effective-one-body Hamiltonian}
\label{appendix}

Following Ref.~\cite{Da.al3.00}, but including also the logarithmic contributions that were left out
of their analysis, we show that it is possible to build EOB Hamiltonians 
such that (i) the mapping (\ref{Hreal}) holds at all PN orders and (ii) the non-geodesic terms 
in the square brackets of Eq.~(\ref{Heff}) are proportional to the radial momentum, at all PN orders, thus vanishing for circular 
orbits.

We start by considering the generic structure of the (specific) two-body 
Hamiltonian in the center-of-mass frame, at a given PN order, as a
function of the reduced canonical variables $\hat{\boldsymbol{p}}=\boldsymbol{p}/\mu$ and $\boldsymbol{q} = \boldsymbol{r}/M$:
\begin{align}\label{eq3.6}
	\widehat{H}_{(n,k){\rm PN}}({\boldsymbol r},{\boldsymbol p}) &= (\ln q)^k \, \Bigg\{\hat{p}^{2(n+1)}  \nonumber \\
	&+ \frac{1}{q} \left[ \hat{p}^{2n}+\hat{p}^{2n-2}(n\hat{p})^2+\cdots+(n\hat{p})^{2n}\right] \nonumber \\
	&+ \frac{1}{q^2} \left[\hat{p}^{2(n-1)}+\cdots+(n\hat{p})^{2(n-1)}\right] \nonumber \\ 
	& + \cdots+ \frac{1}{q^{n+1}}\Bigg\}\,,
\end{align}
where we introduce the notations $q \equiv \sqrt{\boldsymbol{q}\cdot \boldsymbol{q}}$ and $(n\hat{p}) \equiv \boldsymbol{n}\cdot \hat{\boldsymbol{p}}$, and where we use a subscript $(n,k)$PN to denote the contribution to the $n$PN-accurate Hamiltonian which is $\mathcal{O}(c^{-2n})$ and proportional to $(\ln{q})^k$, with $k \geqslant 0$. Indeed, the general structure of the near-zone expansion (formally $r/c \to 0$) of the PN metric is known to be of the type $[\ln{(r/c)}]^k (r/c)^n$ \cite{BlDa.86}, yielding terms proportional to $(\ln{q})^k$ in the conservative dynamics starting at 4PN order \cite{BlDa.88} (see also \cite{Bl.al2.10} for a more recent discussion).

The number of independent coefficients in the Hamiltonian \eqref{eq3.6} is easily found to be \cite{Da.al3.00} 
\begin{equation}
C_{\rm H}(n,k) = \frac{(n+1)\,(n+2)}{2} + 1\,. 
\end{equation}
The mapping of the real Hamiltonian onto the effective Hamiltonian can be achieved through a canonical 
transformation \cite{BuDa.99,Da.al3.00}. The most general generating function reads
\begin{align}
	G_{(n,k){\rm PN}}({\boldsymbol r},{\boldsymbol p}) &= (\ln q)^k \, ({\boldsymbol r}\cdot{\boldsymbol p}) \left\{ \hat{p}^{2n} + \frac{1}{q} \left[ \hat{p}^{2(n-1)}  \right. \right. \nonumber \\ &\qquad \left. \left.+ \cdots + \, (n\hat{p})^{2(n-1)} \right] + \cdots + \frac{1}{q^n} \right\} . 
\end{align}
Thus, the number of independent coefficients in the generating function is \cite{Da.al3.00}
\begin{equation}
C_{\rm G}(n,k) = \frac{n\,(n+1)}{2} + 1\,. 
\end{equation}
At 3PN or higher orders, non-geodesic (NG) terms resulting in quartic or higher powers of the momentum $\hat{\boldsymbol{p}}$ 
can appear inside the square brackets of the effective Hamiltonian (\ref{Heff}). At 3PN order, the NG terms symbolically read
\begin{equation}
{\cal Q}_{3{\rm PN}}({\boldsymbol r},{\boldsymbol p}) = \frac{1}{q^2} \left[ \hat{p}^{4} + \hat{p}^{2}\,(n\hat{p})^2 + (n\hat{p})^4 \right] .
\end{equation}
(The first logarithms appear at 4PN order.) At higher PN orders ($n >3$), the generic structure is 
\begin{align}
	{\cal Q}_{(n,k){\rm PN}}({\boldsymbol r},{\boldsymbol p}) &=  
(\ln q)^k \, \Bigg\{\frac{1}{q^2} \left[ \hat{p}^{2n-2} + \hat{p}^{2n-4}\,(n\hat{p})^2 \nonumber \right.\\ & \left.+\cdots + 
(n\hat{p})^{2n-2} \right] + \cdots + \nonumber \\& \frac{1}{q^{n-1}} \left[ \hat{p}^{4} + \hat{p}^{2}\,(n\hat{p})^2 + (n\hat{p})^4 \right] \Bigg\} \, .
\end{align}
It is straightforward to derive that the number of arbitrary coefficients in ${\cal Q}_{(n,k){\rm PN}}$ is 
\begin{equation}
C_{\rm NG}(n,k) = \frac{(n+3)\,(n-2)}{2}\,,
\end{equation}
hence the number of NG terms that depend on the radial momentum $(n\hat{p})$ is 
\begin{equation}
C^{\rm circ}_{\rm NG}(n,k) = C_{\rm NG}(n,k) - (n-2) = \frac{(n+1)\,(n-2)}{2}\,.
\end{equation}
Given that at the PN order $(n,k)$ we have two new coefficients in the effective metric 
potentials $A$ and $\bar{D}$, multiplied respectively by $u^{n+1} (\ln{u})^k$ and $u^n (\ln{u})^k$, but no new coefficient in the mapping between the real and 
effective Hamiltonians, because we assume that (\ref{Hreal}) is valid at all PN orders, 
we obtain that the difference between the number of equations to satisfy and the number of unknowns is 
\begin{align}
\delta(n,k) &= C_{\rm H}(n,k) - C_{\rm G}(n,k) - 2 - C^{\rm circ}_{\rm NG}(n,k) \nonumber \\ &= - \frac{n\,(n-3)}{2}\,,
\end{align}
which is always zero or negative starting at 3PN ($n=3$). This proves that for all $n \geqslant 3$ (and any $k \geqslant 0$), it is possible to build the mapping between the real and the effective descriptions in such a way that Eq.~(\ref{Hreal}) holds, and circular orbits follow from a ``geodesic'' effective Hamiltonian [\textit{i.e.}, 
all the higher-momentum terms inside the square brackets in Eq.~(\ref{Heff}) can be chosen to be
proportional to the radial momentum $(n\hat{p})$]. It is for this class of EOB Hamiltonians that 
we have determined the effective metric potentials $A$ and $\bar{D}$ at all PN orders, linearly in the symmetric mass ratio $\nu$. 

\bibliography{}

\end{document}